**Nanowire-Intensified MEF in Hybrid Polymer-Plasmonic Electrospun Filaments**

*Andrea Camposeo,\* Radoslaw Jurga, Maria Moffa, Alberto Portone, Francesco Cardarelli, Fabio Della Sala, Cristian Ciracì,\* Dario Pisignano\**

Dr. A. Camposeo, Dr. M. Moffa, Dr. F. Cardarelli, Prof. D. Pisignano,
NEST, Istituto Nanoscienze-CNR,
Piazza San Silvestro 12, I-56127 Pisa, Italy
E-mail: andrea.camposeo@nano.cnr.it

R. Jurga, Dr. F. Della Sala, Dr. C. Ciracì
Center for Biomolecular Nanotechnologies@UNILE,
Istituto Italiano di Tecnologia,
Via Barsanti 14, I-73010 Arnesano, Italy
E-mail: cristian.ciraci@iit.it

Dr. F. Della Sala
Institute for Microelectronics and Microsystems (CNR-IMM),
Via Monteroni, Campus Unisalento, I-73100 Lecce, Italy

R. Jurga, A. Portone
Dipartimento di Matematica e Fisica "Ennio De Giorgi",
Università del Salento,
via Arnesano, I-73100 Lecce, Italy

Prof. D. Pisignano
Dipartimento di Fisica,
Università di Pisa,
Largo B. Pontecorvo 3, I-56127 Pisa, Italy
E-mail: dario.pisignano@unipi.it

The enhancement of the radiative processes in organic and inorganic light-emitting systems has been intensively investigated in the last decades,[1-3] being of interest both for elucidating nanoscale photophysical processes at fundamental level, and for developing applications in the fields of spectroscopy, biodiagnostics, and nanophotonics. Fluorescence enhancement can arise from a cooperative process, such as superradiance found in atoms, quantum wells, and molecular crystals,[4-7] or from the interaction of emitters with dielectric photonic crystals[2] and nanoantennas.[8] In addition, enhanced emission has been evidenced for fluorophores in close proximity of a metal surface or nanostructure.[9] Such metal-enhanced fluorescence (MEF) occurs when a substantial overlap exists for the absorption spectrum of the emitter and the scattering band due to the localized surface plasmon resonance in the metal.[10] Furthermore, MEF requires the emitter-metal separation to be in a well-defined range, in order to prevent fluorescence quenching which takes place at very short distances.[11] MEF has been found with systems emitting in the visible and near-infrared spectral range, and exploited to build biosensors, light-emitting microarrays, and tunable nanolasers.[3,12,13]

However, while metal nanoparticles (NPs) and nanostructures provide large fluorescence enhancement factors and potential confinement of light across sub-wavelength volumes, they have intrinsically high Ohmic losses. These prevent long-range transport of light, which is relevant for routing and coupling different nanoscale light sources in integrated photonic systems. Conversely, dielectric structures are highly transparent and can guide photons over macroscopic distances with limited optical losses. These properties have motivated the interest toward crossbreed systems, in which subwavelength light confinement of surface plasmons and long-range propagation of guided modes of dielectric waveguides are hybridized for developing novel and more efficient nanophotonic platforms.[14,15] For instance, NPs can be used as local antennas for in- and out-coupling light with waveguides positioned nearby.[15-17] Moreover, in subwavelength waveguides the significant intensity gradients associated to light confinement lead to coupling the photon spin and orbital angular momentum, which in turn is found to strongly modify the scattering properties of metal NPs across the nanofiber evanescent field.[17] The interaction of two counter-propagating modes with the localized plasmon resonance of an embedded array of NPs is also found to support transparent waveguide plasmon-polariton modes, through classical electromagnetically induced transparency, leading to slow guided light propagation.[18] Metal NPs positioned on the surface of subwavelength waveguides have also been exploited for developing nanoscale acoustic detectors, based on the response of a nanofiber to incident sound waves modulating the NP-fiber separation.[19] Recently, hybrid lasing systems with optical emissions at subwavelength scale have been demonstrated, by coupling Ag nanowires to dielectric perovskite crystals and organic microdisk cavities.[20,21]

Most of these hybrid systems have been realized by depositing, or lithographically-defining metal NPs and nanostructures on the surface of passive waveguides, with coupling allowed by evanescent field. This configuration, however, might limit applications, due to the high sensitivity of the resulting photonic components to ambient perturbations and the consequently frequent need for operation in vacuum.[17] In this respect, metal particles embedded within the core of active waveguides would lead to much more stable and optically robust systems. Various methods are currently available for the production of functional elongated nanostructures[22-24] and active dielectric filaments embedding NPs,[25,26] including electrospinning which allows continuous and multifunctional composite nanofibers to be obtained. Single photons[27] and ensemble emission[28] coupling to modes of electrospun nanofibers have been reported, as well as fluorescent filaments used as active components in lasers,[29] fluorescent barcodes,[30] and UV excitation sources.[31] Fibers doped with metal NPs have been employed for optical sensing through surface enhanced Raman scattering,[32] for Li ion storage,[33] and for catalysis.[34] Little has been done for developing active organic nanofibers embedding plasmonic NPs, although these devices might feature localized MEF of internal fluorophores,[35] and enable improved nanoscale light sources and lasers.[36,37]

Here we highlight twofold-enhanced fluorescence in hybrid nanowires, namely the novel Nanofiber-Enhanced MEF (NE-MEF) effect, realizing fluorescent organic-plasmonic, electrospun nanowires. These hybrid systems are found to exhibit MEF higher than in thin films. The photonic architecture combines active polymer fibers with subwavelength diameter and metal NPs for local enhancement of the emission of incorporated fluorophores. A broadband increase of the fluorescence enhancement is evidenced for fibers compared to films, which is rationalized by finite-element method (FEM) simulations unveiling the unique features of the field spatial distribution and of the emitter quantum yield in the confined geometry.

Our devices are produced by embedding Au NPs (average size 60 nm, inset of **Figure 1**a) and Rhodamine 6G (R6G) fluorophores in polyvinylpyrrolidone (PVP), here used as nanofiber matrix for its optical transparency at visible wavelengths and excellent fiber-forming properties via electrospinning.[38] The absorption and photoluminescence (PL) spectra of R6G-doped PVP exhibit peaks at ~540 nm and ~565 nm, respectively, whereas Au NPs feature an extinction spectrum with peak wavelength at 538 nm (Figure S1), well matching the optical features of the fluoresphore. Au NPs with smaller (larger) average size having blue (red)-shifted extinction spectra are expected to lead to less efficient fluorescence enhancement mechanisms.[10] The free-standing, hybrid fibers once collected on transmission electron microscopy (TEM) grids, are displayed in Figure 1a,b. The geometry with air surrounding the fibers allows for improving field confinement and light transport through the waveguides. The electrospun fibers have diameter, $d$, down to 200 nm, and a smooth surface (inset of Figure 1b) that is very important to lower scattering losses, thus leading to propagation lengths of the order of 100 μm.[27,28]

Figure 1c shows a scanning transmission electron microscopy (STEM) image of the fibers, highlighting the incorporation of an individual Au NP. The position of NPs is mainly determined by the dynamics of the entangled network of the polymer macromolecules in the jet during the electrospinning process. Hence, a significant role is played by the viscoelastic properties of the electrified solution, as well as by the very high strain rates (e.g., $10^3$-$10^5$ s$^{-1}$) promoted by electrospinning. These might lead to a local increase of the polymer density at the jet center as previously probed by X-ray imaging,[39,40] and to the formation of relatively ordered, anisotropic conformations of both the polymer macromolecules and the embedded particles.[32,41,42] In addition, the density of particles in the fibers can be controlled by varying the molar ratio of NPs to the polymer in the electrospun solution.[32] This strategy allows filaments embedding isolated NPs, or linear chains of NPs, to be achieved.

The PL of individual hybrid fibers are investigated by fluorescence confocal microscopy, allowing both intensity and spectral properties to be mapped with a spatial resolution given by the diffraction-limited excitation spot size. Figure 1d shows the confocal map of the fiber displayed in Figure 1c, clearly evidencing a spot with brighter fluorescence in correspondence of the Au NP. The corresponding fluorescence enhancement factor ($\eta_{fiber}$) is obtained by comparing the PL intensity of the brighter spot ($I_{Au}$) with that of nearby regions where no particle is present ($I_F$). Such combined STEM and confocal analysis, performed on several fibers with 300 nm diameter, leads to an average enhancement factor, $\eta_{fiber}=I_{Au}/I_F$ =1.6 ± 0.1. The role of MEF in this effect is highlighted by time-resolved PL measurements performed with confocal detection,[43] which clearly show a decrease of the luminescence lifetime of the chromophores in correspondence of the Au NP (**Figure 2**a). In fact, while PL time-resolved profiles of nanofiber-embedded chromophores display almost mono-exponential decays with average lifetime about 0.8 ns, the PL decay becomes multi-exponential in presence of Au NPs, with average lifetimes about 0.5 ns.

Furthermore, to unveil the contribution of the filamentous geometry in amplifying the enhancement, namely in leading to NE-MEF, similar measurements are performed on films having identical composition and comparable thickness to the fibers. Interestingly, $\eta_{fiber}$ is found to be higher than the fluorescence enhancement factor measured in films ($\eta_{film}$=1.3±0.1, see Figure S2 for details). Therefore, an additional effect is associated to the nanofiber confined geometry, synergistic with the MEF induced by the metal NPs. Furthermore, this effect is dependent on the fiber transversal size, as well as on the polarization of the excitation laser. Figure 2b-d show the micro-PL (μ-PL) intensity profiles measured along the longitudinal axis of three electrospun fibers with different diameter, evidencing a different MEF enhancement factor in correspondence of a single Au NP, depending on the $d$ value. Decreasing the fiber diameter from 600 to 200 nm leads to an increase of $\eta_{fiber}$ by about 60%. **Figure 3**a displays how the enhancement depends on the polarization of the excitation laser, evidencing improved NE-MEF for excitation light polarized parallel to the fiber length. Instead, the fluorescence enhancement factor is found not to significantly depend on wavelength. The measured PL spectrum of R6G nearby the metal particles in the fibers is shown in Figure 3b, together with a spectrum in a region without NPs. Overall, the hybrid polymer-plasmonic nanostructures provide NE-MEF with an almost constant $\eta_{fiber}$ on a spectral range of about 80 nm (Figure 3c), a property important for those applications where broadband light sources are used, including ultrafast spectroscopy and optical sensing.

In order to better elucidate the NE-MEF process and to discriminate the contribution from the metal NP and from the organic filaments to the ultimate values of $\eta_{fiber}$, classical electromagnetism FEM simulations are carried out for our system. The fluorophore is approximated as a point dipole, assumed to be weakly excited (no saturation) by an external

Gaussian beam impinging orthogonally to the fiber longitudinal axis, as sketched in **Figure 4**a. The calculated spatial distribution of electric field in the hybrid Au NP/fiber is also presented in Figure 4a for a PVP filament with 300 nm diameter (a detailed comparison of the field distribution in and nearby the fiber is performed and shown in Figure S3 for filaments either with or without NPs). A local enhancement up to a factor 3.8 is achieved in close proximity of the Au NP, whereas an increase by a factor about 2 is evident in a region nearby the NP with extension of the order of the beam waist. For comparison, we report in Figure 4b the electric field distribution for the case of an Au NP embedded in a PVP film of 300 nm in thickness. The maximum local enhancement factor for the film is ~3.2 around the NP (~40% lower intensity than in the fiber case). Furthermore, the field remains visibly reduced in the surrounding region as well. Additional field enhancement can be achieved by decreasing the fiber size, as shown in Figure S4, where the spatial distribution of the electric field in hybrid filaments is calculated for fibers with diameters of 200 and 600 nm. By decreasing the fiber size, the electric field nearby the Au NP is enhanced up to a factor 4.2.

The ratio of the radiative decay rate ($\gamma_r$) to the total decay rate ($\gamma_{sp}$), namely the quantum yield of the fluorophores embedded into the fibers, $q = \gamma_r / \gamma_{sp} = (\gamma_{sp} - \gamma_{nr} - \gamma_{int}) / \gamma_{sp}$, is derived by obtaining $\gamma_{int}$, which represents the internal emission rate accounting for intrinsic decay such as phononic or trapped states, from the measured quantum yield of the chromophores in bulk, and by calculating $\gamma_{nr} = \frac{1}{2} \frac{\gamma_r^0}{W_r^0} \int_\Omega \mathrm{Re}\{\mathbf{J}^* \cdot \mathbf{E}\} dV$. The integration of all metal losses is performed over the NP volume ($\Omega$), and $\gamma_{sp}$, from the Green's function, $\mathbf{G}$, of the system:

$$\gamma_{sp} = \frac{2\omega^2}{\hbar\varepsilon_0 c^2} |\mathbf{p}|^2 \left[ \mathbf{n}_p \cdot \mathrm{Im}\{\mathbf{G}(\mathbf{r}_m, \mathbf{r}_m)\} \cdot \mathbf{n}_p \right] + \gamma_{int} \qquad (1)$$

where $\mathbf{p}$ is the transition dipole moment, $\omega$ is given by the transition frequency, $\mathbf{n}_p$ is the unit vector of the dipole moment, which accounts for its orientation, and $\mathbf{G}$ is the Green dyadic function calculated at the position $\mathbf{r}_m$ of the emitting dipole. In Figure 4c-e we show the calculated values of the quantum yield, $q = q_{\mathbf{n}_p}(\mathbf{r}_m)$, for different orientations of the dipole as a function of its position inside the fiber, normalized with respect to the quantum yield in bulk (the corresponding $\gamma_{sp}$ values are shown in Figure S5). We point out that the calculation of full quantum-yield maps as a function of the position and orientation of the emitter is quite challenging, as about one thousand different FEM calculations are required. The field patterns due to both the fiber and the Au NP are clearly visible in Figure 4c-e. In particular, when the emitter is parallel to the NP surface (Figure 4c), the quenching near the metal allows to appreciate the effect of the fiber confinement, which doubles the quantum yield with respect to its bulk value. In the other cases (Figure 4d-e), the NP enhances the quantum yield up to 8 times in the fibers, with a relatively lower effect of fiber confinement. Indeed, such position and orientation dependence of the quantum yield and of the decay rates leads to a complex temporal behavior of the fluorescence following NP addition, as shown in Figure 2a.[44]

The orientation-averaged fluorescence intensity, $PL(\boldsymbol{r})$, is obtained as:

$$PL(\boldsymbol{r}) = \frac{A}{3} \sum_{i=\rho,\phi,z} |E_i(\rho,\phi,z)|^2 q_i(\rho,z) \tag{2}$$

where $q_\rho(\rho,z) = q_y(0,y,z)$ and $q_\phi(\rho,z) = q_x(0,y,z)$, are calculated in the cross-section plane of Figure 4c-e, and $A$ is a constant. Figure 4f shows the resulting map of the fluorescence enhancement with respect to the bulk, for an averaged random distribution of chromophore orientations: enhancements up to 30 times the bulk values can be achieved. Of course such enhancements are larger than those found in experiments, since the various positions and orientations of the fluorophores are averaged over a volume corresponding to the excitation spot size during measurements. In order to account for such features of experiments, we average $PL$ with respect to the volume included in a sphere of diameter $D$ around the NP, as depicted in Figure 4f. To this aim, from the definition in Eq. (2), it is easy to average over a volume $\Omega$, as $\overline{PL} = \frac{1}{\Omega} \int_\Omega PL(\mathbf{r}) \, dV$. The resulting, volume-averaged fluorescence enhancement with respect to the case without Au NP, $\eta_{av} = \overline{PL_{fiber+NP}}/\overline{PL_{fiber}}$ is shown in Figure 4g as a function of $D$, together with the corresponding values calculated for a 300 nm thick film. The effect of the Au NP can be clearly appreciated in a small volume in the vicinity of the NP. Considering the experimental spot used for excitation, a quite good agreement is found with the experimental results shown in Figure 3b, with $\eta_{av}$ around 1.5-1.6. More importantly, $\eta_{av}$ is found to be larger than the value for the film in all the investigated range of $D$. A dependence of $\eta_{av}$ on the polarization of the excitation light is also obtained (Figure S6), in agreement with experimental findings (Figure 3a). Moreover, for emitters oriented perpendicularly to the fiber length a large fraction of the emitted light is coupled to modes waveguided along the organic filament (Figure S7), and the presence of the NP strongly increases the amount of photons available in such modes (Figure 4h).

Such cooperative NE-MEF and improved channeling to guided modes might be exploited for enhancing the coupling efficiency of single-photon emitters embedded in dielectric waveguides.[27] Other applications of these findings embrace coupling metallic particles with nanofiber architectures to achieve active photonic components and integrated systems with higher emission efficiencies, or miniaturized lasers with lower threshold. These include plasmon-enhanced random lasers based on nanofiber networks[37] for optical tagging and fluorescence sensors, and devices combining optical amplifiers based on organic filaments[45] with plasmonic metamaterials.[46] In addition NE-MEF, being inherently localized at subwavelength scales, opens interesting perspectives for imaging, nanolithography and 3D manufacturing,[47,48] possibly leading to a significant improvement in terms of achievable spatial resolution.

In summary, special features for the enhancement of the fluorescence in hybrid polymer nanofibers embedding plasmonic NPs are found and rationalized. The subwavelength fiber architecture leads to NE-MEF, namely to a significant improvement of the MEF compared to the effect observed for plasmonic particles embedded in films. μ-PL measurements indicate a size-dependent effect, which is significantly improved by decreasing the fiber diameter. In addition, NE-MEF is broadband, being almost insensitive to fluorescence wavelength in a range of about 80 nm. These results are supported by numerical simulations, in which the fiber shows a maximum increase of ~40% in PL with respect to an extended film of comparable thickness. These hybrid nanostructures, providing improved local field enhancement are an example of how more efficient and tailored emission can be achieved by cooperatively exploiting metal and dielectric nanostructures, which might lead to enhanced performance in multi-spot light-emitting devices with high efficiency, mutually inter-coupled through nanofiber circuits and networks for integrated information processing.

**Experimental Section**

*Electrospinning and nanofiber morphology*. Hybrid nanofibers were produced by electrospinning a solution of 100 mg of PVP (molecular weight $1.3 \times 10^6$ g/mol, Alpha Aesar) and 1 mg of R6G (Exciton) dissolved in 1 mL mixture of ethanol (0.85 mL) and distilled water (0.15 mL). The used water dispersion contained $5 \times 10^{11}$ Au NPs (Sigma Aldrich), corresponding to a weight ratio of particles with respect to the PVP matrix of about 1:100. Such value is found to allow well-isolated Au NPs to be embedded in the electrospun fibers, as required for optical characterization. Further increase of the weight ratio of Au NPs can be exploited to enhance the macroscopic averaged fluorescence of the hybrid fibers, because of the higher number of spots with intensified fluorescence. The solution was fed by a syringe pump (Harvard Apparatus), with constant flow rate (0.5 mL/h) through a metallic needle (gauge 27), and fibers were collected on a metallic plate positioned at a distance of 15 cm from the spinneret. The voltage bias between the spinneret and the collector was 11 kV. Free-standing fibers were obtained by deposition on TEM grids (TAAB Laboratories Equipment Ltd.), whose supporting structure also provided a reference frame to localize individual fibers to image by scanning electron microscopy (SEM, FEI Nova NanoSEM 450), STEM, and confocal fluorescence microscopy. Fibers were firstly imaged by an optical inverted microscope (Eclipse Ti, Nikon) in bright field, to be precisely labelled with respect to the grid, and then analyzed sequentially by confocal microscopy and STEM. The combination of the optical and morphological data allows the bright spots observed in confocal microscopy to be univocally correlated to incorporated NPs. Reference thin films were spin-cast on quartz substrates (5000 rpm, 60 s).

*Optical measurements*. Absorption measurements were carried out by using a UV-visible spectrophotometer (Lambda 950, Perkin Elmer). PL spectra were obtained by exciting the samples with a diode laser and collecting the emission by an optical fiber, coupled to a spectrometer (USB4000, Ocean Optics). Samples were mounted in an integrating sphere (Labsphere), which allows the emission quantum yield to be measured (3% for R6G molecules embedded in PVP). A μ-PL analysis was performed by using a confocal microscopy system, based on an inverted microscope (Eclipse Ti, Nikon) equipped with a confocal laser scanning head (A1 MP, Nikon). The samples were excited by an Ar$^+$ laser ($\lambda_{exc}$=488 nm) through a 20× objective (numerical aperture, *NA*=0.5), which was used also to collect the sample emission, and analyzed by means of spectral detection unit with a multi-anode photomultiplier. To account for the finite collection angle, the ratio $\eta_{fiber}=I_{Au}/I_F$ was considered in the analysis, which does not depend on the numerical aperture of the collection optics as shown in Figure S8. The spot size of the focalized excitation laser beam was about 300 nm. The intensity of the excitation laser transmitted through the sample was measured by a photomultiplier, simultaneously to the fluorescence signal. Scattering images of thin films were obtained by illuminating the samples at almost grazing incidence with a white light, fiber-coupled Tungsten lamp, while collecting the light diffused by the samples by the microscope objective used for confocal fluorescence measurements. Confocal fluorescence lifetime imaging microscopy (FLIM) was performed by an inverted microscope with confocal head (TCS SP5, Leica Microsystem) and a 20× objective (*NA*=0.5). A 470 nm pulsed diode laser operating at 40 MHz was used as excitation source, and the fluorescence intensity measured by a photomultiplier tube interfaced with a Time Correlated Single Photon Counting setup (PicoHarp 300, PicoQuant, Berlin). FLIM acquisitions lasted until an average of $10^2$-$10^3$ photons were collected in each pixel.

*Simulations.* FEM full-wave simulations were performed by using COMSOL. The nanofiber was modeled as a dielectric cylinder of radius *R*=150 nm, length *L*=2 μm and permittivity ε=2.37 (the PVP experimental value) containing an Au NP at its center and surrounded by air. Perfectly matched layers (PMLs) around the geometry have been used in order to emulate an infinitely extended geometry. The fluorophores were modeled as a monochromatic point dipole emitting at the wavelength λ=565 nm (corresponding to the measured peak emission wavelength, as shown in Figure S1). The reference quantum yield $q_{bulk}$ was set to the measured experimental value (3%) in bulk PVP. The Green function **G(r,r)** was evaluated by varying the position of the dipole emitter on a discrete 50×100 grid placed on the positive quadrant of the *y-z* plane at the center of the fiber and taking advantage from the axis symmetry of the system with respect to *z*, and from the mirror symmetry with respect to the *x-y* plane. We point out that each point of such maps required a separated three-dimensional FEM calculation. The excitation field was modeled by imposing a spatially dependent surface current profile on a spherical simulation boundary below the PML of radius 700 nm. The profile was approximated using a Gaussian beam polarized along the fiber axis (*z*) with waist $w_0 = 300$ nm, focused 200 nm below the center of the fiber. Note that although the Gaussian beam approximation does not hold for such a small beam, this approach allows to easily define a converging beam. Analogous methods were used for the simulations of films. In this case, a quartz substrate of index $n = 1.46$ has been considered in agreement with experimental conditions.

**Acknowledgements**
The research leading to these results has received funding from the European Research Council under the European Union's Seventh Framework Programme (FP/2007-2013)/ERC Grant Agreement n. 306357 (ERC Starting Grant "NANO-JETS"). The Apulia Networks of Public Research Laboratories WAFITECH (09) and M. I. T. T. (13) are also acknowledged.

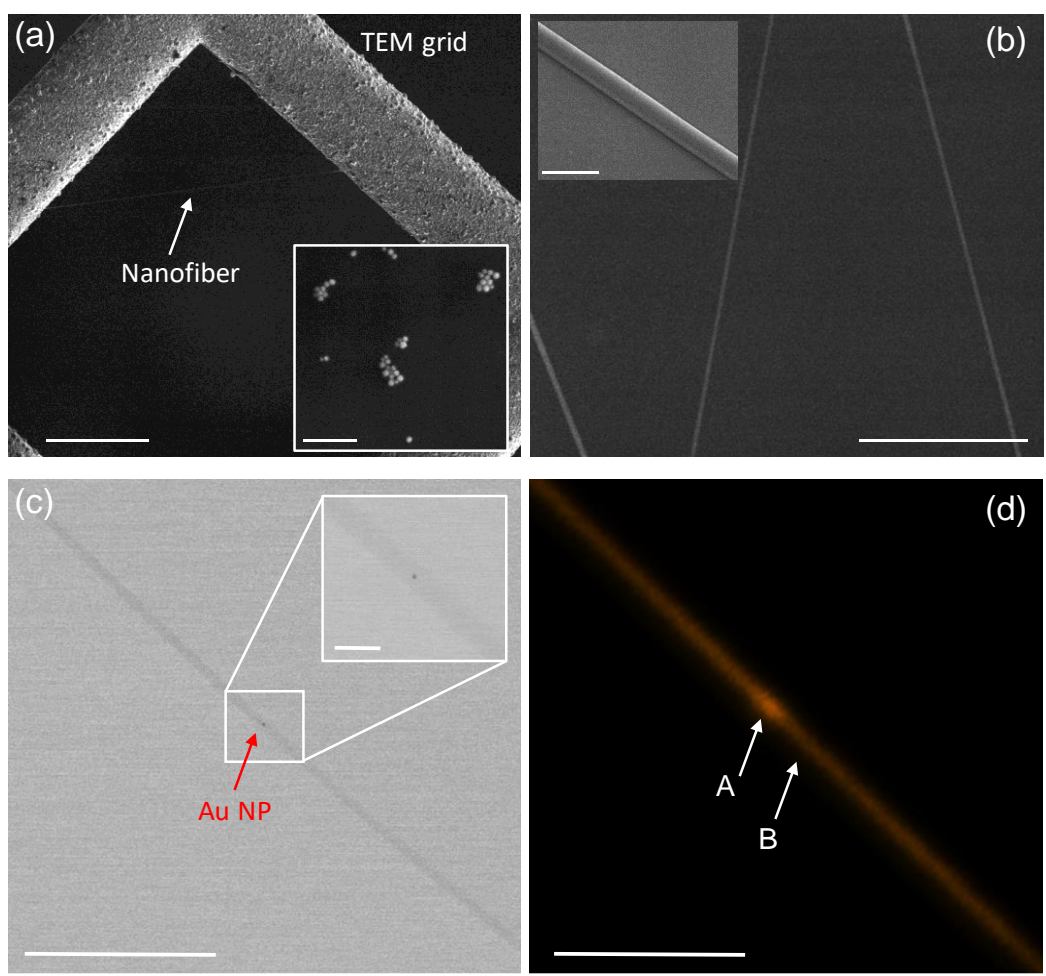

**Figure 1.** (a)-(b) SEM micrographs of free-standing R6G-doped PVP nanofibers [highlighted by the arrow in (a)], embedding Au NPs. The fibers are supported by a grid for TEM. Scale bars: 50 µm (a) and 10 µm (b), respectively. Insets: Au NPs on a Si substrate (a, scale bar: 500 nm) and individual fiber at high magnification (b, scale bar: 1 µm). (c) STEM micrograph of a fiber embedding a single Au NP (arrow). Scale bar: 5 µm. Inset: zoom of the region with the NP. Scale bar: 500 nm. (d) Confocal map of the fluorescence intensity collected from the fiber shown in (c). Scale bar: 5 µm. The arrow labelled as 'A' highlights the fiber region embedding the Au NP, whereas the arrow labelled as 'B' points a region without particles.

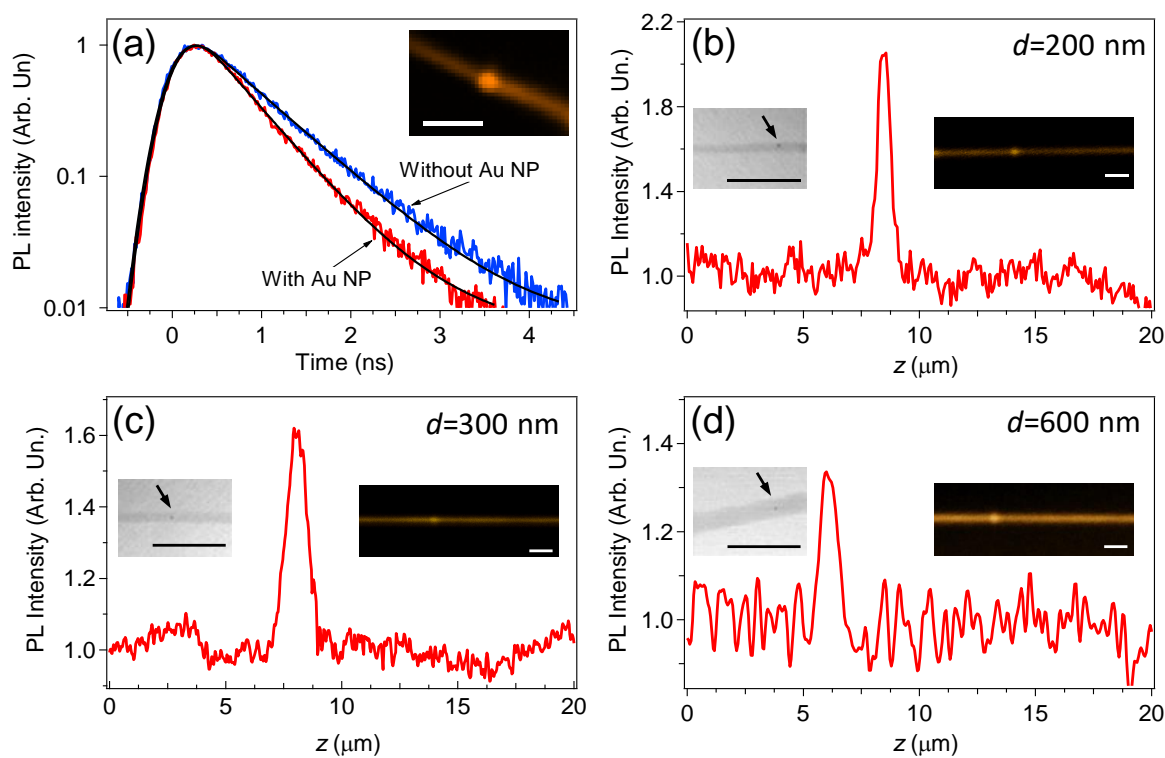

**Figure 2.** (a) Temporal decay curves for PL intensity, measured in correspondence of the Au NP embedded in the nanofiber (red lines) and in a region without Au NPs (blue line). The continuous lines are fit to the data by exponential functions convoluted with the instrumental response function. Inset: corresponding confocal micrograph. Scale bar: 5 μm. (b-d) Fiber size-dependence of the fluorescence enhancement. PL intensity profiles vs. longitudinal position ($z$) along the fiber axis, as obtained by confocal microscopy. Fiber diameter ($d$): 200 nm (b), 300 nm (c), and 600 nm (d), respectively. Detection wavelength range: 560-580 nm. Right insets: corresponding confocal fluorescence micrographs. Scale bars: 2 μm. Left insets: corresponding STEM micrographs highlighting the hybrid PVP fiber and incorporated Au NP (arrows). Scale bars: 2 μm.

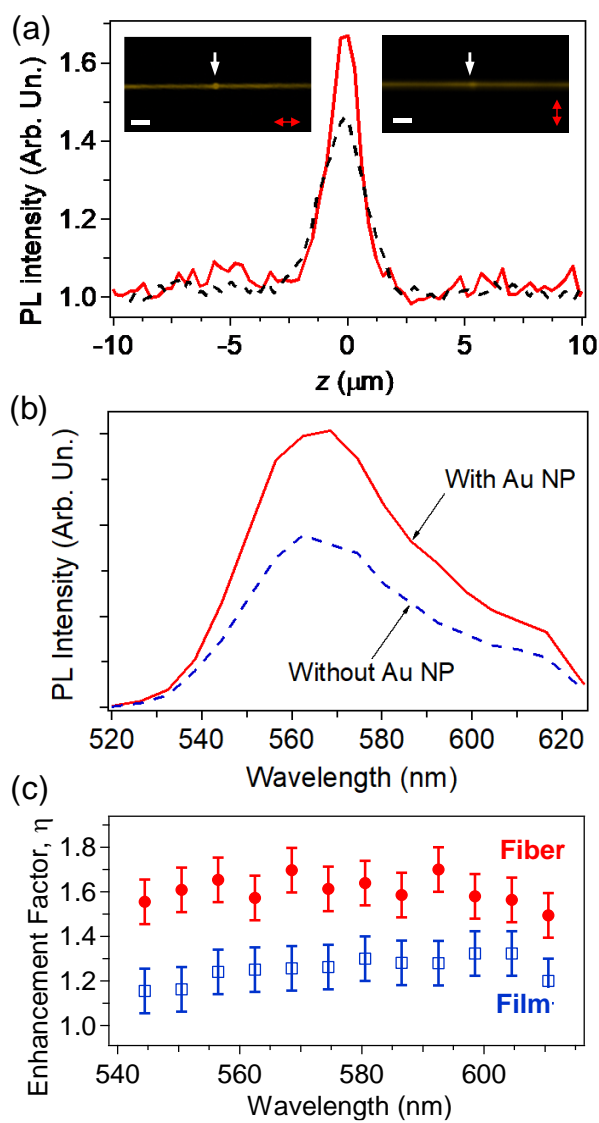

**Figure 3.** (a) PL intensity profiles vs. longitudinal position (z) along the fiber longitudinal axis, for polarization of the excitation laser parallel (continuous line) or perpendicular (dashed line) to the fiber length. Insets: corresponding confocal fluorescence micrographs. The embedded Au NPs are highlighted by vertical white arrows, whereas the red arrows indicate the direction of polarization of the excitation laser. Scale bars: 2 μm. (b) μ-PL spectrum measured in correspondence of the Au NPs embedded in the nanofiber (continuous line, 'A' region in Figure 1d). The μ-PL spectrum measured in a region without NPs (dashed line, 'B' region in Figure 1d) is also displayed. These spectra are obtained by integrating the PL intensity collected from a fiber area of about 1 μm². (c) Spectral dependence of the fluorescence enhancement factor for fibers (full circles) and for thin films (empty squares). Fiber diameter = 300 nm.

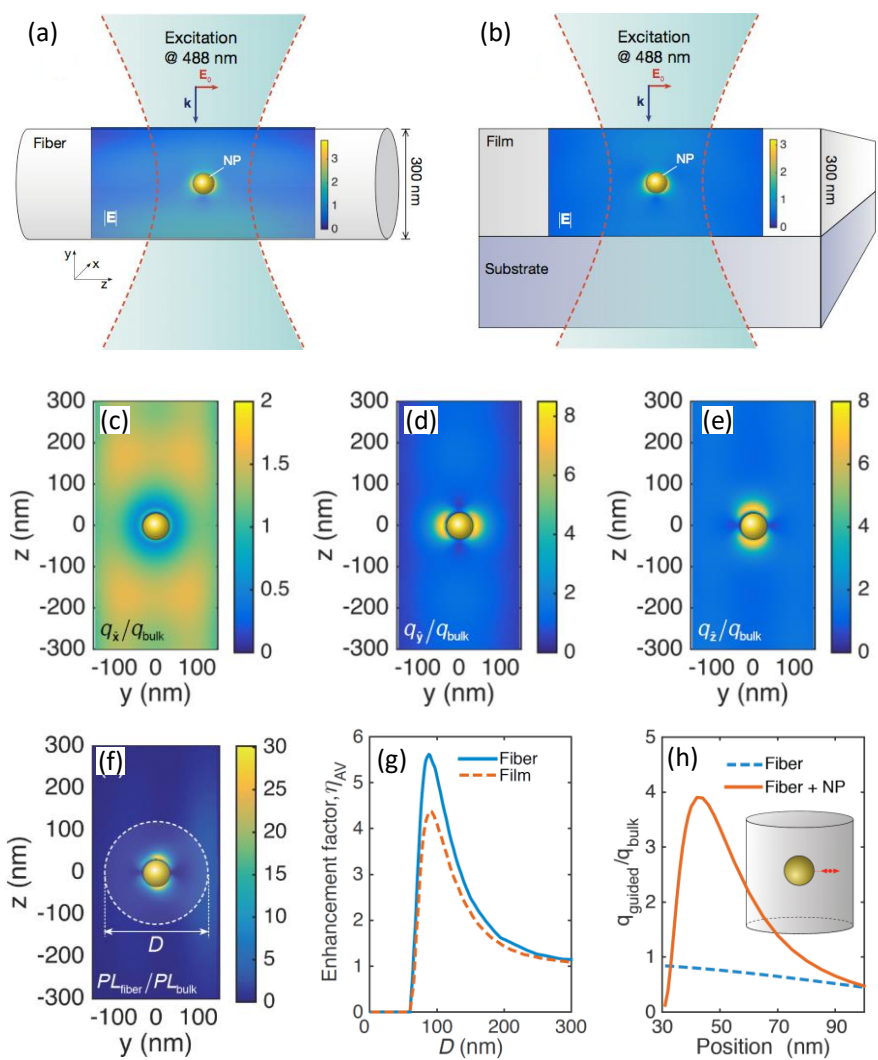

**Figure 4.** (a)-(b) Schematics of the hybrid Au NP/fiber (a) and the Au NP/film (b) systems. The particle is positioned at the center of the fiber (film). The area containing the NP is excited by a beam focused at its center. The dashed lines sketch the spatial variation of the excitation beam waist. The map shows the spatial distribution of the electric field, $\mathbf{E}$, normalized to the maximum filed of the incident beam in vacuum. (c)-(e) Quantum yield map, as a function of the position of the emitter inside the fiber and for different orientations of the emitting dipole. The maps are calculated considering the dipole unit vector $\mathbf{n}_p$, parallel to the $x$ (c), $y$ (d), and $z$ (e) axis, respectively. The data are normalized to the reference quantum yield, $q_{bulk}$, measured for R6G-doped in PVP (see Experimental Section). Fiber diameter: 300 nm. (f) Fluorescence enhancement with respect to the bare bulk, for a random distribution of chromophore orientations. The dashed circle highlights the volume used for averaging the fluorescence intensity in (g). (g) Volume-averaged enhancement of the fluorescence intensity of a hybrid system (fiber+Au NP) with respect to a pristine PVP fiber (i.e. a fiber without Au NPs), vs. spherical domain diameter ($D$) depicted in (f) (continuous line). The same curve for the volume-averaged enhancement in film is also reported (dashed line). (h) Quantum yield of guided fluorescence for a dipole perpendicular to the fiber length, as a function of its distance from the Au NP as depicted in the inset. Here, $q_{guided}$ is given by the number of guided photons per photon absorbed by the emitter.

# Supporting Information

**Nanowire-Intensified MEF in Hybrid Polymer-Plasmonic Electrospun Filaments**

*Andrea Camposeo,\* Radoslaw Jurga, Maria Moffa, Alberto Portone, Francesco Cardarelli*
*Fabio Della Sala, Cristian Ciracì,\* Dario Pisignano\**

Dr. A. Camposeo, Dr. M. Moffa, Dr. F. Cardarelli, Prof. D. Pisignano,
NEST, Istituto Nanoscienze-CNR,
Piazza San Silvestro 12, I-56127 Pisa, Italy
E-mail: andrea.camposeo@nano.cnr.it

R. Jurga, Dr. F. Della Sala, Dr. C. Ciracì
Center for Biomolecular Nanotechnologies@UNILE,
Istituto Italiano di Tecnologia,
Via Barsanti 14, I-73010 Arnesano, Italy
E-mail: cristian.ciraci@iit.it

Dr. F. Della Sala
Institute for Microelectronics and Microsystems (CNR-IMM),
Via Monteroni, Campus Unisalento, I-73100 Lecce, Italy

R. Jurga, A. Portone
Dipartimento di Matematica e Fisica "Ennio De Giorgi",
Università del Salento,
via Arnesano, I-73100 Lecce, Italy

Prof. D. Pisignano
Dipartimento di Fisica,
Università di Pisa,
Largo B. Pontecorvo 3, I-56127 Pisa, Italy
E-mail: dario.pisignano@unipi.it

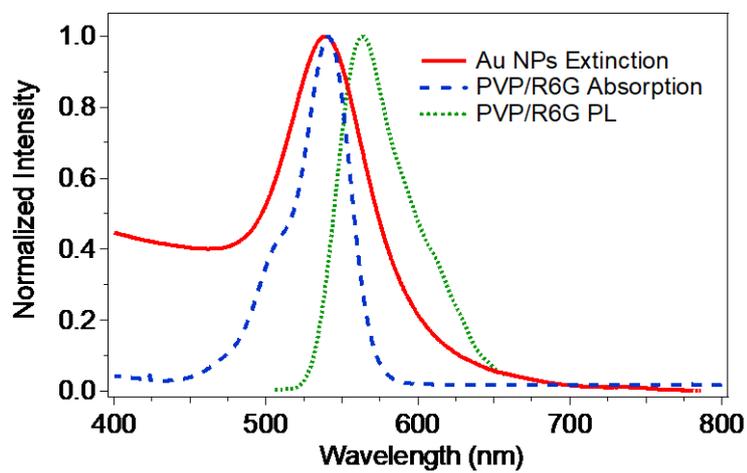

**Figure S1**. Extinction spectrum of Au NPs dispersed in water (continuous line). Absorption (dashed line) and PL (dotted line) spectra of a R6G-doped PVP film.

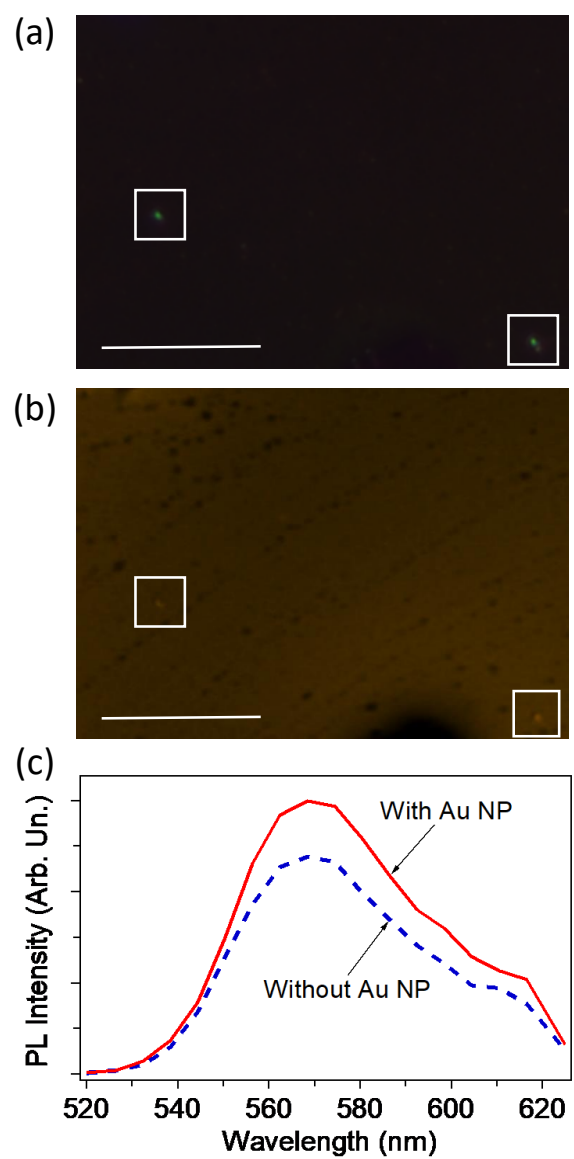

**Figure S2**. (a) Optical scattering and (b) corresponding confocal fluorescence micrographs of a R6G-doped film embedding Au NPs. Scale bars: 20 μm. In (a) NPs are visible as bright green spots (highlighted by square boxes), due to light diffusion at wavelengths around 538 nm (extinction spectrum in Figure S1). The PL intensity micrograph (b) shows enhanced emission where particles are present (square boxes). (c) μ-PL spectrum obtained by confocal analysis, with (continuous line) and without Au NP (dashed line).

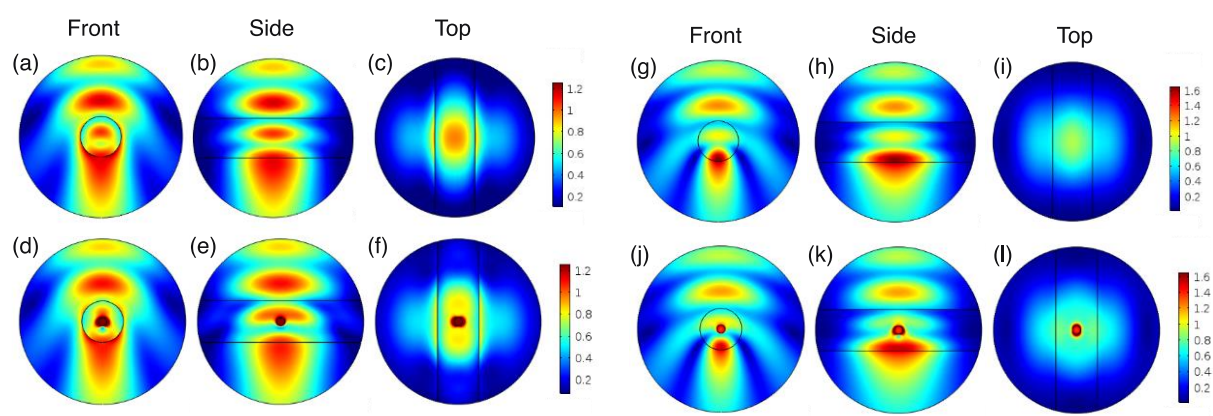

**Figure S3.** Effect of the dielectric on the electric field enhancement, |**E**| / $E_0$, for a nanofiber without Au NP (top panels) and with the addition of the NP (bottom panels). The nanofiber surface is highlighted as black continous line. The incident field is a beam polarized normally to the fiber axis in panels (a-f) and parallel to the fiber axis in (g-l). Three different cross-sectional views are reported: front (plane $xy$), side (plane $yz$) and top (plane $xz$). $x$,$y$ and $z$ axes are defined in Figure 4a.

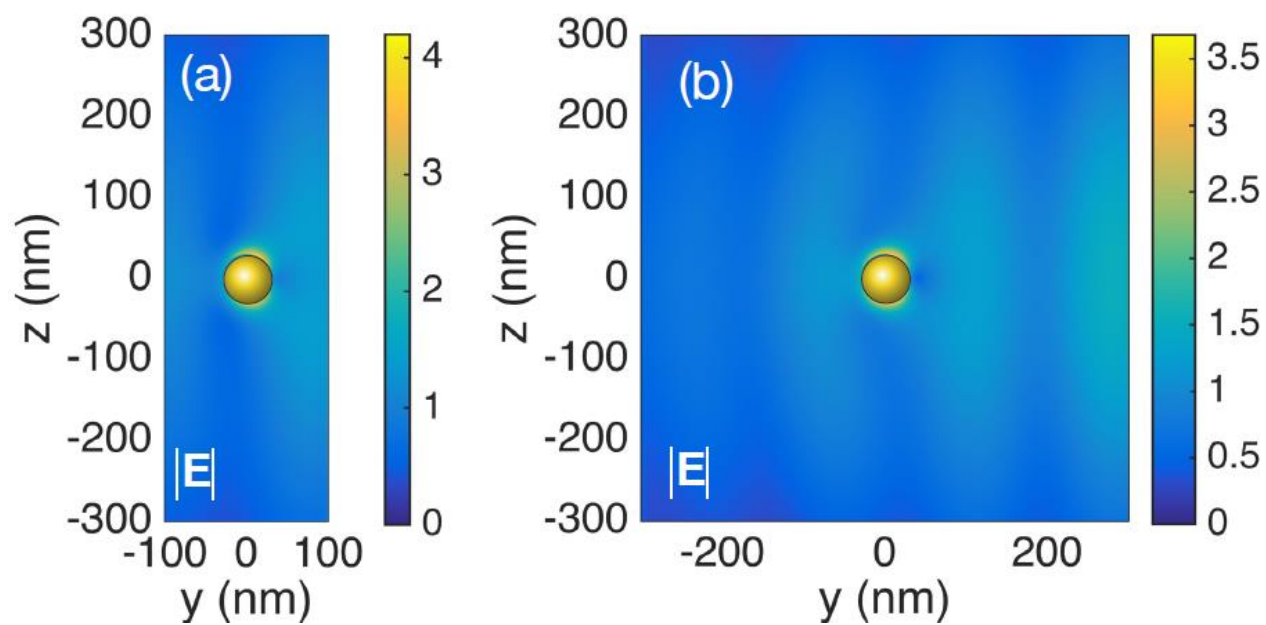

**Figure S4**. (a)-(b) Maps of the spatial distribution of the electric field, **E**, normalized to the maximum filed of the incident beam in vacuum for fibers with a diameter of 200 and 600 nm, respectively.

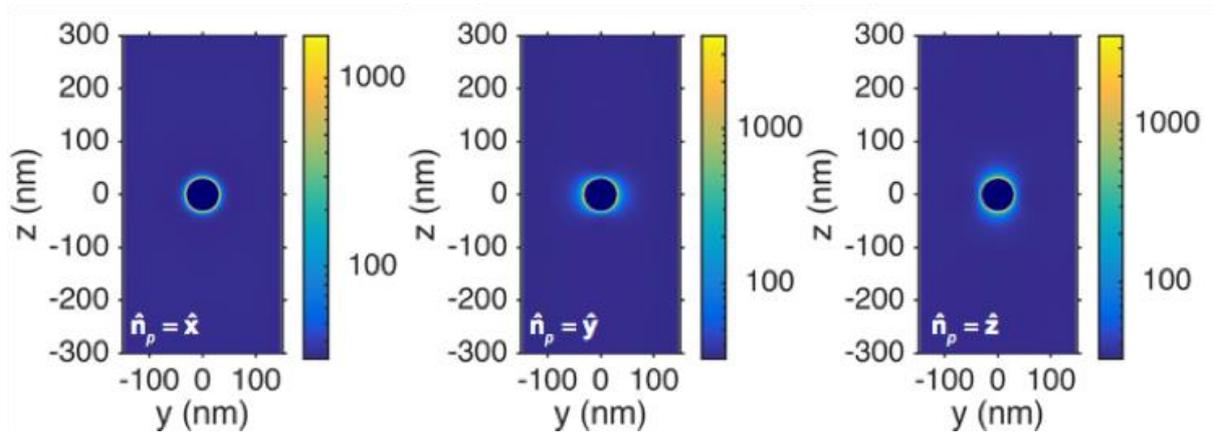

**Figure S5.** Total emission rate ($\gamma_{sp}/\gamma_r^0$) as a function of the position of the emitter inside the fiber and for different orientations of the dipole $\boldsymbol{n}_p$ . The emission rates are normalized with respect to the radiative emission rate in free space $\gamma_r^0 = \frac{\omega^3 |\mathbf{p}|^2}{3\hbar\pi\epsilon_0 c^3}$ .

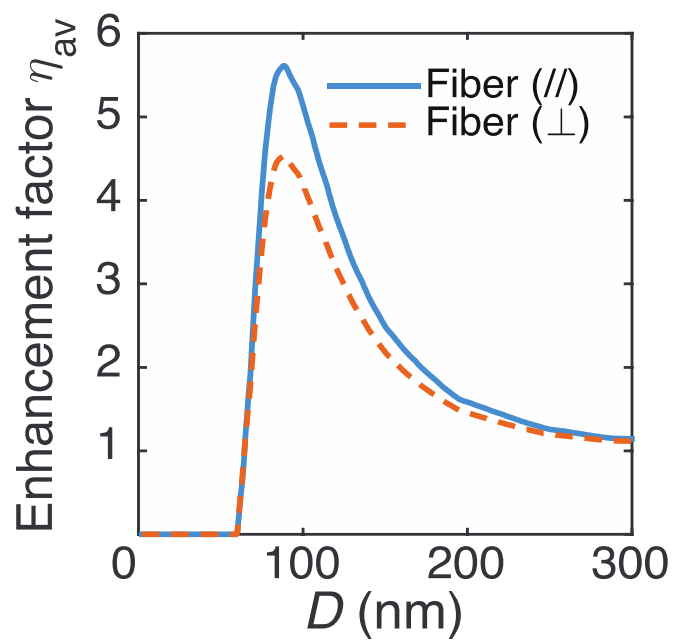

**Figure S6.** Averaged photoluminescence enhancement for polarization of the incident field parallel and perpendicular to fiber axis.

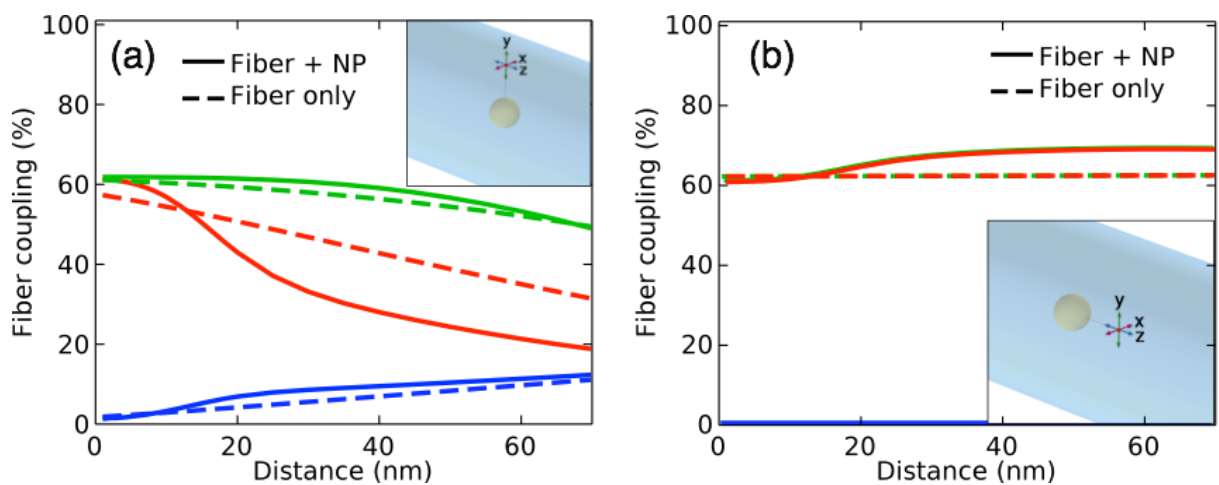

**Figure S7.** Coupling of the dipole energy into the fiber (guided) as a function of the distance of the dipole from the NP inside the fiber, for different dipole orientations (as shown in the insets: $\mathbf{n}_p = \boldsymbol{x}, \boldsymbol{y}, \boldsymbol{z}$ in red, green and blue respectively). In (a) the dipole is displaced along the $y$ direction. In (b) the dipole is displaced along the $z$ direction: note that for the $\mathbf{n}_p = \boldsymbol{z}$ there is no coupling, while because of the cylindrical symmetry the case of $\mathbf{n}_p = \boldsymbol{x}$ is exactly the same as $\mathbf{n}_p = \boldsymbol{y}$. The coupling coefficients are calculated by projecting the emitted field on the fiber modes via the scalar product: $\int_S E \times H^* \cdot \mathbf{n}_S dS$, with $S$ the cross-section area and $\mathbf{n}_S$ the unit vector normal to $S$.

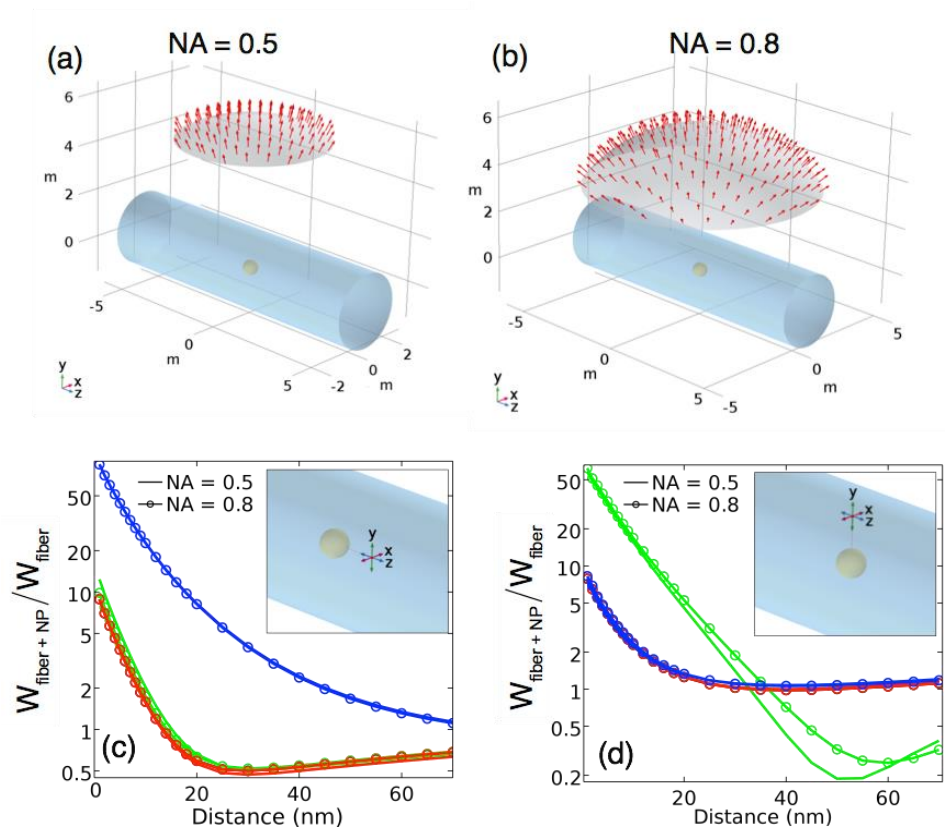

**Figure S8.** Influence of the numerical aperture (*NA*) on measurements. (a)-(b) *NA* in the simulation domain; the solid angle corresponds to *NA*=0.5 in (a) and *NA*=0.8 in (b). (c)-(d) Emitted power through the *NA* for different dipole orientations (as shown in the insets: $\mathbf{n}_p = \boldsymbol{x}$, $\boldsymbol{y}$, $\boldsymbol{z}$ in red, green and blue respectively), as a function of the dipole position and normalized with respect to the power measured in the case of fiber only. In (c) the dipole is displaced along the *y* direction. In (d) the dipole is displaced along the *z* direction. Note that the ordinate axis is in log scale.